\documentclass[prl,aps,twocolumn,showpacs,amsmath,amssymb]{revtex4}
\usepackage{graphicx}
\usepackage{dcolumn}
\usepackage{bm}
\usepackage{amsmath}
\usepackage{amssymb}
\newcommand{\Rmnum}[1]{\expandafter\@slowromancap\romannumeral #1@}
\newcommand{\rmnone}{\rm I}
\newcommand{\rmntwo}{\rm II}
\newcommand{\Tk}{T_{\rm K}}
\newcommand{\Lp}{{L+}}
\newcommand{\Lm}{{L-}}
\newcommand{\Rp}{{R+}}
\newcommand{\Rm}{{R-}}
\newcommand{\Tpp}{\vert\! +\!+\rangle}
\newcommand{\Tmm}{\vert\! -\!-\rangle}
\newcommand{\Tzero}{\vert T0 \rangle}
\newcommand{\Singlet}{\vert S \rangle}
\newcommand{\chicloc}{\chi_\text{c}^\text{loc}}
\newcommand{\Hflip}{H_{\rm flip}^{\rm eff}}

\newcommand{\HCO}{H_{\rm CO}^{\rm eff}}
\newcommand{\HCOt}{H_{\rm COt}^{\rm eff}}
\newcommand{\Ukt}{U_\text{KT}'}

\makeatletter

\begin{document}

\title{Quantum Phase Transition and Dynamically Enhanced Symmetry in\\
Quadruple Quantum Dot System}

\author{Dong E. Liu, Shailesh Chandrasekharan, and Harold U. Baranger}
\affiliation{
Department of Physics, Duke University, Box 90305, 
Durham, North Carolina 27708-0305, USA
}
\date{August 5, 2010}

\begin{abstract}
We propose a system of four quantum dots designed to study the competition
between three types of interactions: Heisenberg, Kondo and Ising. We find a rich
phase diagram containing two sharp features: a quantum phase transition (QPT)
between charge-ordered and charge-liquid phases, and a dramatic resonance in the
charge liquid visible in the conductance. The QPT is of the Kosterlitz-Thouless
type with a discontinuous jump in the conductance at the transition. We connect
the resonance phenomenon with the degeneracy of three levels in the isolated
quadruple dot and argue that this leads to a Kondo-like dynamical enhancement of
symmetry from $U(1)\!\times\! Z_2$ to $U(1)\!\times\!  U(1)$. 
\end{abstract}

% 73.23.Hk 	Coulomb blockade; single-electron tunneling 
% 73.21.La 	Quantum dots
% 72.10.Fk 	Scattering by point defects, dislocations, surfaces, 
%               and other imperfections (including Kondo effect) 
% 05.30.Rt 	Quantum phase transitions
% 11.30.Na 	Nonlinear and dynamical symmetries (spectrum-generating symms)
% 11.30.Qc 	Spontaneous and radiative symmetry breaking

\maketitle

Strong electronic correlations create a variety of interesting phenomena
including quantum phase transitions \cite{Sachdev09}, emergence of new
symmetries \cite{ColdeaE810}, non-Fermi-liquid behavior
\cite{StewartRMP01,ChangRMP03}, and Kondo resonances \cite{ColemanRev02}. It is
likely that new, yet undiscovered, phenomena can arise from unexplored competing
interactions. Today, quantum dots provide controlled and tunable experimental
quantum systems to study strong correlation effects. Further, unlike most
materials, quantum dots can be modeled using impurity models that can be treated
theoretically much more easily. Single quantum dots have been studied
extensively, both theoretically and experimentally, which has led to a firm
understanding of their Kondo physics \cite{GoldhaberGRev07,ChangRev09}. More
recently, the focus has shifted to multiple quantum dot systems where a richer
variety of quantum phenomena become accessible 
\cite{GoldhaberGRev07,ChangRev09}. These include non-Fermi liquids
\cite{Potok06}, dynamical enhancement of symmetry \cite{Kuzmenko0204}, and
quantum phase transitions \cite{vojta06,garst04, Galpin0506}.
% connected to spontaneous symmetry breaking 

In this work we propose a quadruple quantum dot system, that is experimentally
realizable, in which three competing interactions determine the low temperature
physics: (1) Kondo-like coupling of each dot with its lead, (2) Heisenberg
coupling between the dots, and (3) Ising coupling between the dots. Thus, there
are two dimensionless parameters with which to tune the competition. The
pairwise competing interactions, Kondo/Heisenberg and Kondo/Ising, have both
been studied previously. The two impurity Kondo model with a Heisenberg
interaction %, $J_{\rm ex}$, 
between the impurities shows an impurity QPT from separate Kondo screening of
the two spins at small exchange to a local spin singlet (LSS) phase at large
exchange.
% ($J_{\rm ex} > 2.2\,\Tk$, where $\Tk$ is the 
% single impurity Kondo temperature). 
This has received extensive theoretical \cite{vojta06,affleck95, zarand06} 
and experimental \cite{ChangRev09} attention. The competition between Kondo and
Ising couplings has also been studied theoretically for two impurities
\cite{garst04}, including in the quantum dot context \cite{garst04, Galpin0506};
however, no experimentally possible realization of this competition has been
proposed to date.

\begin{figure}[b]
\centering
\includegraphics[width=1.6in,clip]{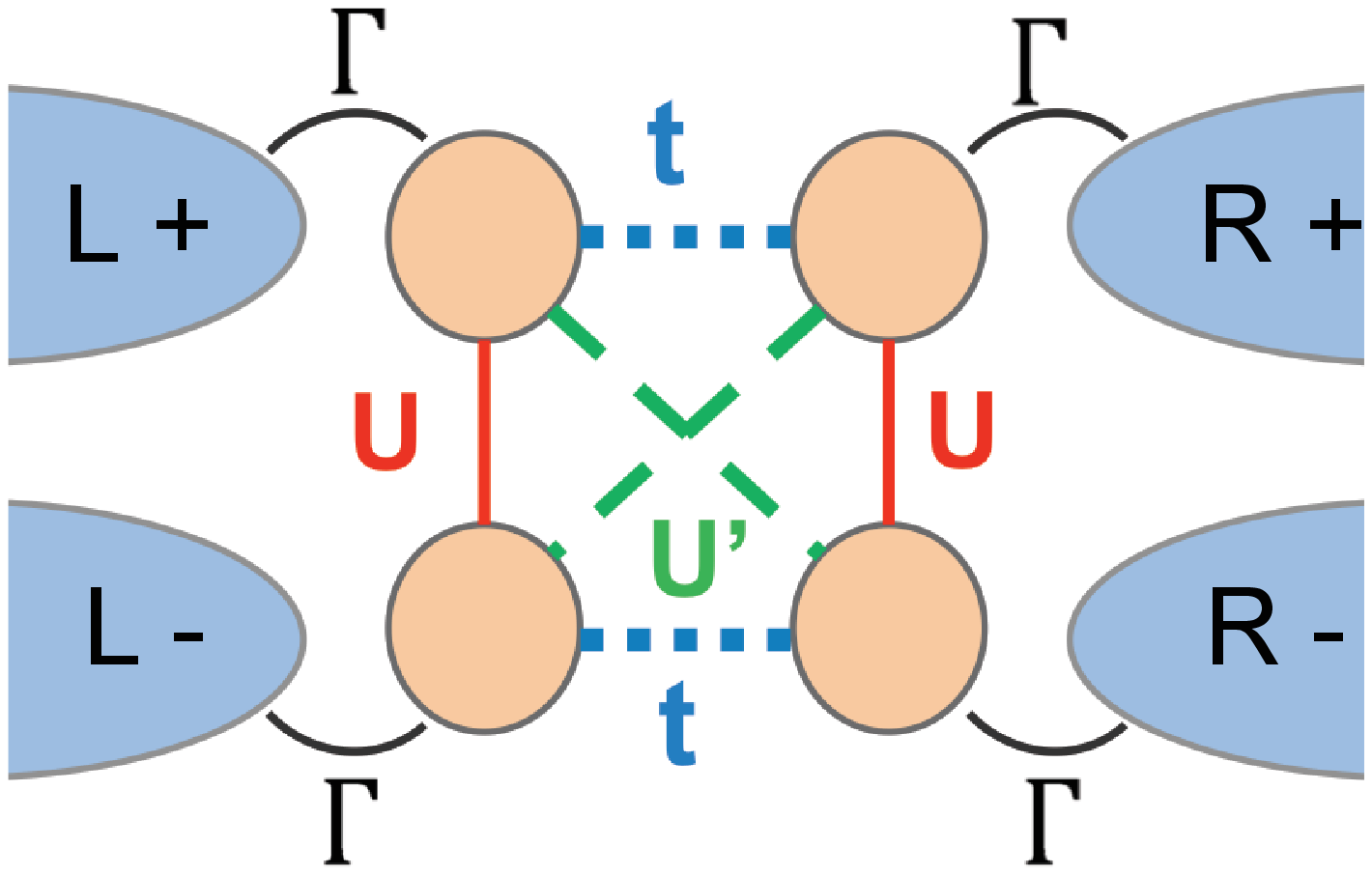}
\includegraphics[width=2.8in]{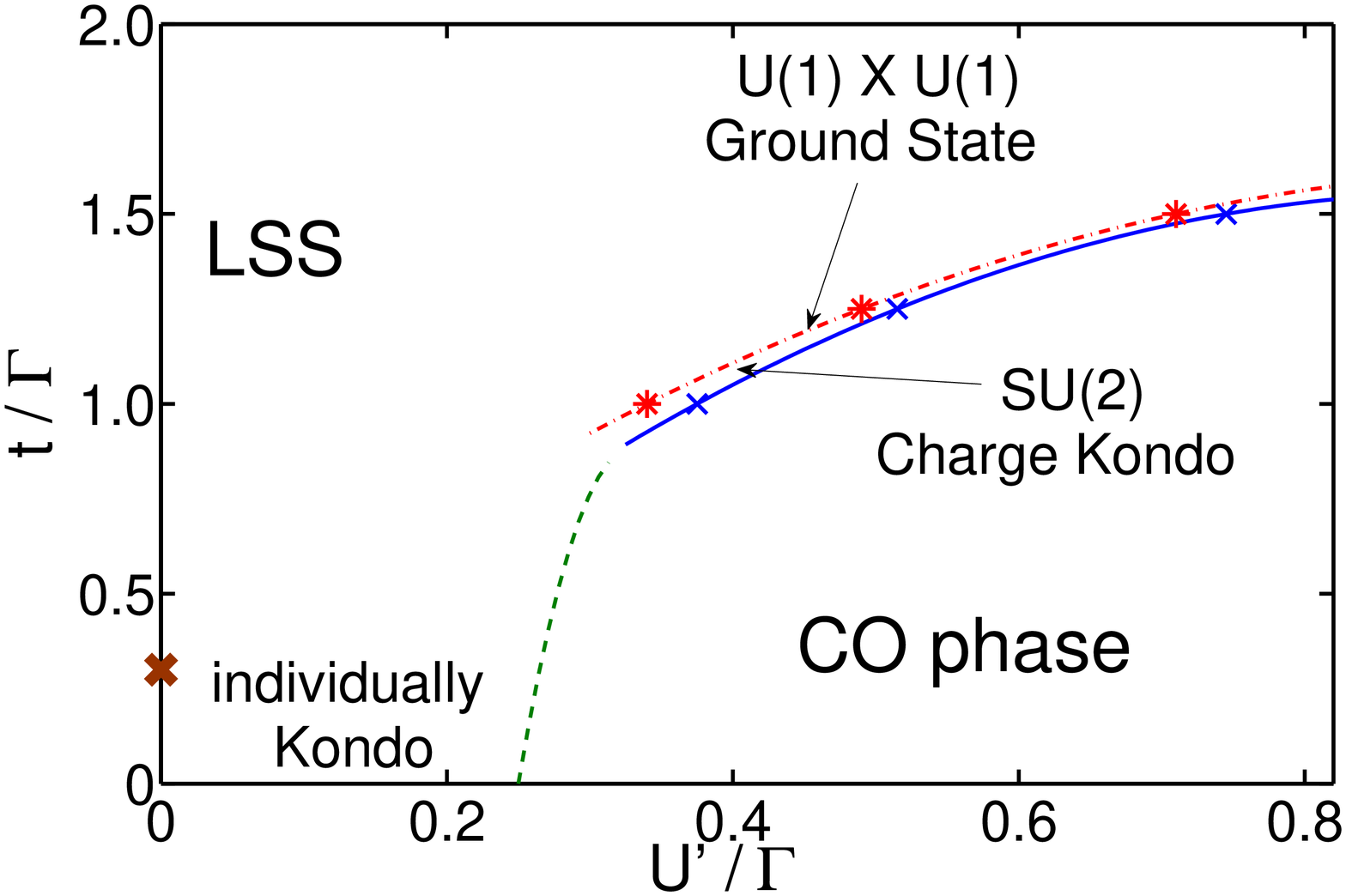}
\vspace*{-0.1in}
\caption{(color online) (a) Quadruple-dot system. $U$ and $U'$ are electrostatic
interactions while $t$ and $\Gamma$ involve electron tunneling. (b)~Ground state
phase diagram as a function of the Ising/Kondo tuning $U'/\Gamma$ and the
Heisenberg/Kondo tuning $t/\Gamma$. The two distinct phases---charge ordered
(CO) and charge liquid---are separated by a KT quantum phase transition [blue
crosses (numerical) and green dashed line (schematic)]. Several cross-overs lie
within the charge-liquid phase. Red stars mark the level crossing where the 
$U(1) \!\times U(1)$ state is found (numerical). The charge Kondo region lies
between the red and blue lines. ``LSS'' denotes the local spin singlet state
(Heisenberg coupling dominates), while when both Heisenberg and Ising couplings
are weak, the system consists of inidividually screened Kondo states on the left
and right.}
\label{fig:phase}
\end{figure}

Our system consists of four quantum dots and four leads, as shown in
Fig.\,\ref{fig:phase}(a), with two polarized (spinless) electrons on the four
dots. We find that the system has a rich phase diagram,
Fig.\,\ref{fig:phase}(b), in terms of the strength of the Heisenberg interaction
controlled by $t$ and the Ising interaction controlled by $U'$. In the absence
of the Ising interaction we start in the local spin singlet (LSS) phase. Upon
increasing the Ising strength, we find that the system first evolves
continuously to a new Kondo-type state with a novel  $U(1)\!\times U(1)$ strong
coupling fixed point. Then there is a crossover to a $SU(2)$ charge Kondo state.
Finally, an additional small increase in $U'$ causes a QPT of the
Kosterlitz-Thouless (KT) type  to a charge ordered state (CO) (as in
Refs.\,\onlinecite{garst04,Galpin0506}) consisting of an unscreened doubly
degenerate ground state \cite{FN_supp}.

{\em Model---}The quantum dots in Fig.\,\ref{fig:phase}(a) are capacitively
coupled in two ways: $U$ is the vertical interaction (between $\Lp$ and $\Lm$;
$\Rp$ and $\Rm$) and $U'$ is along the diagonal (between $\Lp$ and $\Rm$; $\Lm$
and $\Rp$). Along the horizontal, there is no capacitive coupling but there is
direct tunneling $t$ (between $\Lp$ and $\Rp$; $\Lm$ and $\Rm$). Each dot
couples to a conduction lead through $\Gamma=\pi V^{2} \rho$ where $\rho$ is the
density of states of the leads at the Fermi energy. The whole system is
spinless. We consider only the regime in which the four dots contain 2
electrons. 

The system  Hamiltonian is 
$H=H_\text{lead}+H_\text{imp}+H_\text{coup}$, where 
$H_\text{lead}=\sum_{i,s,k} \epsilon_k c_{isk}^{\dagger}c_{isk}$ describes the
four conduction leads ($i=L,R$; $s=+,-$), and 
$H_\text{coup} = V\sum_{i,s,k} (c_{isk}^{\dagger} d_{is} + {\rm H.c.})$
describes the coupling of the leads to the dots which produces the
\textit{Kondo} interaction. $H_\text{imp}$ is the Anderson-type Hamiltonian
\begin{eqnarray}
\label{eq:basic_imp}
\lefteqn{
 H_\text{imp} = \sum_{i=L,R} \sum_{s=+,-} \epsilon_d d_{is}^{\dagger}d_{is} 
 + \sum_{i=L,R} U\hat{n}_{i+}\hat{n}_{i-} 
 } & &   \\
 & & +\; U'( \hat{n}_{L+}\hat{n}_{R-}+\hat{n}_{L-}\hat{n}_{R+} ) 
   + t\!\! \sum_{s=+,-}\! ( d_{Ls}^{\dagger} d_{Rs}+ d_{Rs}^{\dagger} d_{Ls})
  \nonumber
\end{eqnarray}
We take $U \gg U'$ so that there is one electron on the left, and one on the
right. 

We can reformulate $H_\text{imp}$ as an exchange Hamiltonian by noticing that
the right-hand (left-hand) sites form a pseudo-spin: 
$\vec{S}_i=\sum_{s,s'} d_{si}^{\dagger}\vec{\sigma}_{ss'}d_{si}/2$. 
% (i=L,R;s,$s'=+,-$). 
When $t\ll U$, the effective Hamiltonian for the quantum dots is
\begin{equation}
\label{eq:native}
H_\text{imp}^\text{eff} \simeq  
J_{\rm H} \vec{S}_L \cdot \vec{S}_R 
- \widetilde{J}_z S_{L}^z S_{R}^z 
\end{equation}
where $J_{\rm H} \simeq 4t^{2}/(U-U'/2)$ and $\widetilde{J}_z\simeq 2U'$. Thus
$t$ controls the strength of the \textit{Heisenberg} interaction among the dots,
and $U'$ controls the \textit{Ising} coupling. The eigenstates of the impurity
site are the usual (pseudo)spin singlet and triplet states, $\Singlet$, $\Tpp$,
$\Tmm$, and $\Tzero$.
%$\Singlet = 1/\sqrt{2}(d_{L+}^{\dagger} d_{R-}^{\dagger} + 
%d_{R+}^{\dagger} d_{L-}^{\dagger}) \vert 00\rangle$, 
%$\Tpp = d_{R+}^{\dagger} d_{L+}^{\dagger} \vert  00\rangle$, 
%$\Tmm = d_{R-}^{\dagger} d_{L-}^{\dagger} \vert  00\rangle$, and 
%$\vert T0\rangle = 1/\sqrt{2}(d_{L+}^{\dagger} d_{R-}^{\dagger} - 
%d_{R+}^{\dagger} d_{L-}^{\dagger}) \vert 00\rangle$.

Two limits of our model have been studied previously. First, for $U'=0$, it
becomes the well-known two impurity Kondo model \cite{affleck95,zarand06}. If
direct charge transfer is totally suppressed, a QPT occurs between a Kondo
screened state (in which the impurities fluctuate between all four states,
singlet and triplet) and a local spin singlet (LSS) \cite{affleck95,zarand06}.
When direct tunneling is introduced, the QPT is replaced by a smooth
crossover\cite{zarand06}. Second, when $t=0$, the model has
\cite{garst04,Galpin0506} a Kosterlitz-Thouless (KT) type QPT between the Kondo
screened phase at small $U'$ and a charge ordered (CO) phase at large $U'$. The
CO phase has an unscreened doubly degenerate ground state corresponding to
$\Tpp$ and $\Tmm$.  

We solve the model (\ref{eq:basic_imp}) exactly by using finite-temp\-erature
world line quantum Monte Carlo (QMC) simulation with directed loop updates
\cite{Syljuasen02,Yoo05}. We study the regime in which there is a LSS state in
the absence of Ising coupling: $4t^2/U>\Tk^{L/R}$ where $\Tk^{L/R}$ is the Kondo
temperature of the left or right pseudospin individually. Taking the leads to
have a symmetric constant density of states, $\rho = 1/2D$, with half-band-width
$D=2$, we focus on the case $U=3$, $\Gamma=0.2$, and $t=0.3$. $\beta=1/T$ is the
inverse temperature. As $U'$ is varied [a horizontal scan in
Fig.\,\ref{fig:phase}(b)], the gate potential %(not shown) 
is chosen such that $\epsilon_d = -(U+U')/2$, placing the dots right at the
midpoint of the two electron regime.

\begin{figure}[t]
\centering
\includegraphics[width=3.5in,clip]{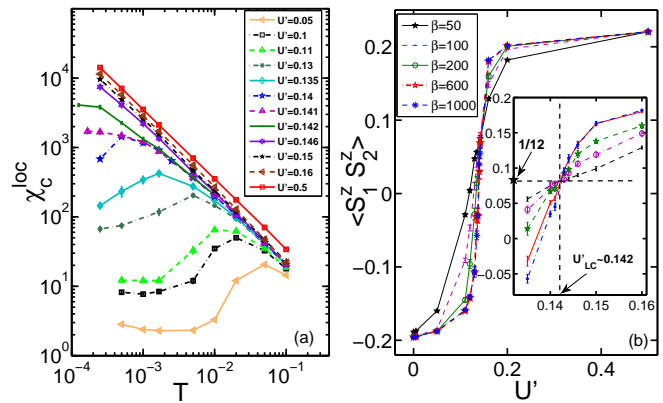}
\vspace*{-0.2in}
\caption{(color online) (a) Local charge susceptibility as a function of
temperature. The power-law behavior of the top three curves indicates the CO
phase. The peak and low-$T$ constant in the lowest curves indicate the LSS
state. The low-$T$ saturation of the middle curves is due to Kondo-like
screening.
(b)~Pesudospin-pseudospin correlation as a function of $U'$ for different
$\beta$. Inset: Zoom near the crossing point. The crossing of the singlet and
doublet levels occurs at $U_{\rm LC}'= 0.142$. corresponding to level crossing.
($U=3$, $\Gamma=0.2$, and $t=0.3$.)
}
\label{fig:stag}
\end{figure}

{\em Thermodynamics---}As a first step toward distinguishing the different
phases, we look at the local charge susceptibility $ \chicloc \equiv
\int_{0}^{\beta} \langle A(\tau) A(0) \rangle d\tau $, where $ A \equiv
n_{L+}+n_{R+}-n_{L-}-n_{R-} $ and $n_{i,s}$ is the charge density of the dot
labeled $i,s$. This could be measured experimentally using recently developed
single charge sensing techniques \cite{Elzerman04}. Fig.\,\ref{fig:stag}(a)
shows $\chicloc$ as a function of temperature for different values of $U'$. The
curves show three types of behavior.
First, for small Ising coupling ($U' \le 0.11$), $\chicloc$ is roughly constant
at low $T$ and has a peak at higher temperature. This is the LSS phase. The
value of $T$ at which $ \chicloc$ peaks decreases as the energy spacing between
the singlet $\Singlet$ and doublet, $\{\Tpp,\Tmm\}$, decreases. 
Second, at the other extreme, for large Ising coupling ($U'\ge 0.15$),
$\chicloc$ behaves as $1/T$ down to our lowest $T$. This is a clear signature of
the CO phase in which the two charge states $\Tpp$ and $\Tmm$ are degenerate.
The tunneling $t$ does not lead to any relevant operator which might destroy the
CO phase \cite{FN_supp}; thus, although the phase boundary is a function of $t$,
the essential nature of the KT QPT [Fig.\,\ref{fig:phase}(b)] is not affected.
Third, for intermediate values of $U'$, $\chicloc$ becomes large and then either
decreases slightly at our lowest $T$ or saturates. This behavior can be produced
by either a near degeneracy between the singlet and doublet states or by charge
Kondo screening of the doublet $\{\Tpp,\Tmm\}$. As we will see from the
conductance data below, the QPT to the CO phase occurs at a value $U_{\rm KT}'$
between $0.146$ and $0.15$.

To extract the position of the level crossing between $\Singlet$ and
$\{\Tpp,\Tmm\}$, we calculate the pseudospin correlation function  $\langle
S_{L}^{z} S_{R}^{z} \rangle $ as a function of $U'$ for different $T$
[Fig.\,\ref{fig:stag}(b)], where $ S_{i}^{z}=(\hat{n}_{i+}-\hat{n}_{i-})/2$. For
$U'=0$, the ground state is the LSS so that $\langle S_{L}^{z} S_{R}^{z} \rangle
\simeq -0.2$ is close to $-1/4$. On the other hand, for large $U'$, in the CO
phase, $\langle S_{L}^{z} S_{R}^{z} \rangle$ is positive and approaches $1/4$.
(The charge fluctuations due to tunneling to the leads causes the values to
differ slightly from $\pm 1/4$.) The crossing point of the curves for different
temperatures gives the position of the (renormalized) level crossing. The inset
shows that it occurs at $\langle S_{L}^{z} S_{R}^{z}\rangle \approx 1/12 $,
which is consistent with the isolated-dots limit. The position of the level
crossing is, then, $U'_{\rm LC}\approx 0.142$; note that this does \textit{not}
coincide with the QPT to the CO phase ($0.146< U_{\rm KT}' <0.15$). 

{\em Conductance---}Conductance is a crucial observable experimentally. However,
QMC is only able to provide numerical data for the imaginary time Green function
at discrete Matsubara frequencies---the conductance cannot be directly
calculated. The zero bias conductance for an impurity model can be obtained
\cite{Syljuasen07} by extrapolating to zero frequency: $G=\lim_{\omega_n \to
0}g(i \omega_n)$ where $g(i \omega_n) \equiv \int_{0}^{\beta} d\tau
\cos(\omega_n \tau)\langle P_x(\tau) P_y(0) \rangle \omega_n/\hbar$ and
$P_y=\sum_{y'\geq y}\hat{n}_{y'}$. We have recently shown that this method works
very well for Anderson-type impurity models in the Kondo region at low
temperature \cite{dong10}. 

\begin{figure}[t]
\centering
\includegraphics[width=3.3in,clip]{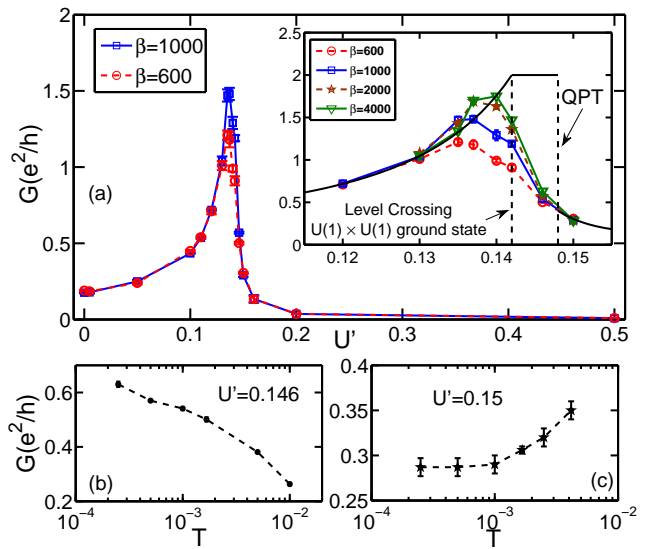}
\caption{(color online) (a) Zero bias conductance as a function of $U'$ for two
values of $\beta$. Inset: Zoom on the peak caused by the $U(1) \!\times U(1)$
ground state. The $T=0$ expectation from the effective theory near the level
crossing is indicated schematically by the black solid line; the two points of
discontinuity (the level crossing and the KT QPT) are marked by dashed lines.
(b),(c)~Conductance as a function of temperature for $U'=0.146$ and $0.15$,
respectively; the opposite trend in these two curves shows that they are on
opposite sides of the QPT.
}
\label{fig:cond}
\end{figure}

We use this method \cite{FN_supp} to find the conductance between the left and
right leads as a function of $U'$ for different $T$; the results are shown in
Fig.\,\ref{fig:cond}. For $U'$ small ($U' \lesssim 0.1$), the conductance is
small because the phase shift is nearly zero in the LSS state \cite{affleck95}.
For $U'$ large ($U' > 0.15$), the conductance is also small and approaches zero
as $U' \to \infty$, consistent with the argument in Ref.\,\onlinecite{garst04}.
At intermediate values of $U'$, there is a strikingly sharp conductance peak
near the value of $U'$ where the level crossing occurs. Here, the conductance
increases as $T$ decreases and approaches the unitary limit $2e^2/h$ as
$T\rightarrow 0$. The position of the conductance peak approaches the level
crossing $U'=0.142$ at low temperature \cite{FN_supp}. Its association with the
level crossing suggests that this peak comes from fluctuations produced by the
degeneracy of $\Singlet$ and $\{\Tpp,\Tmm\}$. 

A sharp jump appears after the peak: notice that the conductance at $U'=0.146$
\emph{increases} at lower temperature while that at $0.15$ \emph{decreases} [see
panels (b) and (c) for clarity]. The latter behavior is the signature of the CO
phase, while the former suggests a Kondo-like phase, namely the dynamic
screening of the $\{\Tpp,\Tmm\}$ doublet. Thus, this sharp jump is associated
with the KT QPT from the screened to the CO phase \cite{garst04}, which occurs
between $U'=0.146$ and $0.15$.

{\em Effective theory near the level crossing---}To gain insight into the
conductance peak, we develop an effective theory near the level crossing. Using
$\Gamma/U$ as a small parameter, we make a Schrieffer-Wolff transformation to
integrate out $\Tzero$; to include tunneling, processes of order $\Gamma t/U^2$
must be included \cite{FN_SWhigherorder}. Higher-order terms in $\Gamma/U$ are
neglected. In the leads, only the combinations $\sum_{k}c_{isk}^{\dagger} \equiv
c_{0,is}^{\dagger}$ need be considered as these are the locations to which the
dots couple. The resulting effective Kondo Hamiltonian reads
\begin{eqnarray}
\label{eq:heff}
H_{\rm Kondo}^{\rm eff} & = & 
  J^{\rmnone}_{\perp}(M_+^{\rmnone} S_{-}^{\rmnone} 
    + M_-^{\rmnone} S_{+}^{\rmnone}) 
    + 2J^{\rmnone}_{z}M_z^{\rmnone} S_{z}^{\rmnone} \nonumber \\
& +&J^{\rmntwo}_{\perp}(M_+^{\rmntwo} S_{-}^{\rmntwo} 
    + M_-^{\rmntwo} S_{+}^{\rmntwo}) 
    + 2J^{\rmntwo}_{z} M_z^{\rmntwo} S_{z}^{\rmntwo} .\;\;\;\;
\end{eqnarray}
The operators $M$ act on the dots,
\begin{eqnarray}
\label{eq:Moperator}
M_+^{\rmnone/\rmntwo} & = & 
\sqrt{2}(\Tpp \langle S \vert \,\mp\, \Singlet \langle -\! -\! \vert) 
= (M_-^{\rmnone/\rmntwo})^\dagger \nonumber \\
M_z^{\rmnone} & = & 
\Tpp \langle +\! +\!  \vert \,-\, \Tmm \langle -\! -\!  \vert \nonumber\\
M_z^{\rmntwo} & = & 
\Tpp \langle +\! +\!  \vert \,+\, 
  \Tmm \langle -\! -\!  \vert - 2\Singlet\langle S\vert
\;,
\end{eqnarray}
while the operators $S$ act on the lead sites,
\begin{eqnarray}
S_{\pm}^{\rmnone} & = & 
 (c_{0,L\pm}^{\dagger}c_{0,L\mp}-c_{0,R\pm}^{\dagger}c_{0,R\mp}) \nonumber\\
S_{z}^{\rmnone} & = & 
 \frac{1}{2}\sum_{i=L,R}(c_{0,i+}^{\dagger}c_{0,i+}-c_{0,i-}^{\dagger}c_{0,i-})
 \nonumber\\
S_{\pm}^{\rmntwo} & = & (\pm c_{0,R\pm}^{\dagger}c_{0,L\mp} \mp
c_{0,L\pm}^{\dagger}c_{0,R\mp})\nonumber\\
S_{z}^{\rmntwo} & = & 
\frac{1}{2}\displaystyle\sum_{s=+,-}(c_{0,Ls}^+c_{0,Rs}+c_{0,Rs}^+c_{0,Ls}) \;.
\label{eq:Soperator}
\end{eqnarray}
For $t/U\ll 1$ and particle-hole symmetry, 
$J_{\rmnone}^{\perp}\simeq J_{\rmnone}^{z}\simeq 4V^2/(U+U')$ 
and $J_{\rmntwo}^{\perp}\simeq J_{\rmntwo}^{z}\simeq 8V^2 t/(U+U')^2$.

Renormalization effects in $H_{\rm Kondo}^{\rm eff}$ can be analyzed using poor
man's scaling \cite{Anderson70}, yielding the scaling equations
\begin{eqnarray}
\label{eq:scaling}
dJ_{\perp}^{\rmnone}/d\ln\! D & = & 
  -2 \rho (J_{\perp}^{\rmnone} J_{z}^{\rmnone} + 
   3 J_{\perp}^{\rmntwo} J_{z}^{\rmntwo}) \nonumber\\ 
dJ_{\perp}^{\rmntwo}/d\ln\! D & = & 
  -2 \rho (J_{\perp}^{\rmntwo} J_{z}^{\rmnone} + 
   3 J_{\perp}^{\rmnone} J_{z}^{\rmntwo})\nonumber\\ 
dJ_{z}^{\rmnone}/d\ln\! D & = & 
  -2 \rho [(J_{\perp}^{\rmnone})^{2} +  
   (J_{\perp}^{\rmntwo})^{2}] \nonumber\\
dJ_{z}^{\rmntwo}/d\ln\! D & = & -4 \rho J_{\perp}^{\rmnone} J_{\perp}^{\rmntwo}
\;.
\end{eqnarray}
Numerical solution of these equations reveals that at a certain value of $D$,
all the coupling constants simultaneously diverge. This defines the problem's
characteristic energy scale $D_0$, which can be considered the Kondo
temperature. The coupling constants have a fixed ratio as they diverge:
$\lim_{D\rightarrow D_0}J_{\perp}^{\rmnone} \!:\! 
J_{\perp}^{\rmntwo} \!:\!  J_{z}^{\rmnone} \!:\! J_{z}^{\rmntwo}
\rightarrow  \sqrt{2}  \!:\!  \sqrt{2}  \!:\!   1  \!:\!  1$, suggesting
dynamical symmetry enhancement of the ground state.

{\em Symmetry analysis---}The six $S$ operators form an $SO(4)$ algebra
\cite{FN_supp}. However, the six $M$ operators do not; rather they form part of
an $SU(3)$ algebra---the missing operators are $\Tpp\langle - -\! |$ and
$\Tmm\langle + +\! |$ \cite{FN_supp}. Since $H_{\rm Kondo}^{\rm eff}\;$ is the
product of two objects which generate different algebras, the symmetry of the
system must be a subgroup of both $SO(4)$ and $SU(3)$. To study the complete
symmetry group of both the bare and fixed-point Hamiltonians, consider the total
$z$-component of pseudospins type $\rmnone$ and $\rmntwo$
\begin{equation}
S_{z,\rm Tot}^{\rmnone} \equiv M_{z}^{\rmnone}+ \sum_{k}S_{z,k}^{\rmnone},\;\;
S_{z,\rm Tot}^{\rmntwo} \equiv M_{z}^{\rmntwo}+ \sum_{k}S_{z,k}^{\rmntwo}
\end{equation}
where $S_{z,k}^{\rmnone/\rmntwo}$ is defined by replacing $c_0$ with $c_k$ in
Eq.\,(\ref{eq:Soperator}). 
One can check that $[S_{z,\rm Tot}^{\rmnone}, H_{\rm lead}+H_{\rm Kondo}^{\rm
eff}] = 0$, which gives a (pseudo)spin $U(1)$ symmetry for the bare Hamiltonian.
The bare Hamiltonian also commutes with interchanging $L$ and $R$ or
interchanging $+$ and $-$. Thus,the symmetry of the bare Hamiltonian is $U(1)_S
\!\times Z_{2,LR} \!\times Z_{2,+-}$ [an irrelevant charge $U(1)$ is ignored]. 

At the fixed point, $\lim_{D\rightarrow D_0}
J_{\perp}^{\rmnone}/J_{\perp}^{\rmntwo}\rightarrow 1$ implies that \emph{both}
$S_{z,\rm Tot}^{\rmnone}$ and $S_{z,\rm Tot}^{\rmntwo}$ commute with the
Hamiltonian. Thus there is an additional $U(1)$ symmetry. Note that $\exp
(i\theta S_{z,\rm Tot}^{\rmntwo})$ generates the $L\leftrightarrow R$
transformation for $\theta=\pi$. Therefore,  the $Z_{2,LR}$ symmetry of the bare
Hamiltonian is dynamically enhanced to a $U(1)$ symmetry. The complete symmetry
group at the fixed point (ground state) is $U(1)_S \times U(1)_{LR} \times
Z_{2,+-}$, where the $Z_{2,+-}$ symmetry is irrelevant for the Kondo physics.

{\em Experimental accessibility---}Though our analysis above is for a highly
symmetric model, possible experimental observation is greatly aided by the fact
that this symmetry is \textit{not} essential. For our scenario, it is crucial
that $\Tpp$ and $\Tmm$ be degenerate, which can be achieved by fine tuning the
gate voltages. When the levels are only approximately degenerate
$\epsilon_{L+}\simeq\epsilon_{L-}\simeq\epsilon_{R+}\simeq\epsilon_{R-}$ (define
$\delta\epsilon$ as the detuning from the symmetric point), $\delta\epsilon$
induces a relevant perturbation in the CO phase \cite{Galpin0506}. However, a
continuous but sharp crossover does still occur in conductance as long as
$\delta\epsilon\ll T_k^{LC}$ and $T\geq\delta\epsilon$ \cite{Galpin0506}, where
$T_k^{LC}$ is the Kondo temperature at the level crossing. 
% The strong coupling fixed point remains stable in this situation. 
Other deviations from symmetry---those involving the tunneling between the dots
or the coupling to the leads, for instance---merely change the coupling
constants in the effective Hamiltonian. In both cases, the $U(1) \!\times U(1)$
strong-coupling fixed point remains stable. Finally, note that the transition
also appears by tuning the experimentally more accessible $t$ at fixed $U'$
\cite{FN_supp} [i.e.\ a vertical cut in Fig.\,\ref{fig:phase}(b)].

This work was supported in part by the 
U.S.\,NSF Grant No.\,DMR-0506953.

\vspace{-0.2in}
%\bibliography{quadruple,cond_qmc,footnotes}

\newpage
\widetext

% \title{Supplementary Material for ``Quantum Phase Transition and \\Dynamically Enhanced Symmetry in Quadruple Quantum Dot System''}
% 
% \author{Dong E. Liu, Shailesh Chandrasekharan, and Harold U. Baranger}
% \affiliation{
% Department of Physics, Duke University, Box 90305, Durham, North Carolina 27708-0305, USA
% }
% \date{\today}
% 
% \maketitle

\begin{center}
\large\bf Supplementary Material for ``Quantum Phase Transition and \\Dynamically Enhanced Symmetry in Quadruple Quantum Dot System''
\end{center}

In this supplementary material we address
(1) extracting the conductance, % from the QMC calculations, 
(2) the shape of the conductance peak, % near the resonance, 
(3) more detail about %the operators in 
    the effective theory near the level crossing, and
(4) an effective theory near the QPT.

\section{I. Conductance from Quantum Monte Carlo}

\begin{figure}[b]
\centering
\includegraphics[width=3.0in]{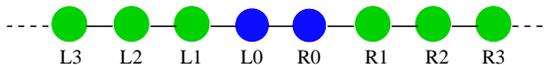}
\vspace*{-0.1in}
\caption{(color online) The 1D infinite tight-binding chain, where $L0$ and $R0$
mark the impurity (quantum dot) sites.
}
\label{fig:chain}
\end{figure}

\begin{figure}[b]
\centering
\includegraphics[width=4.4in,clip]{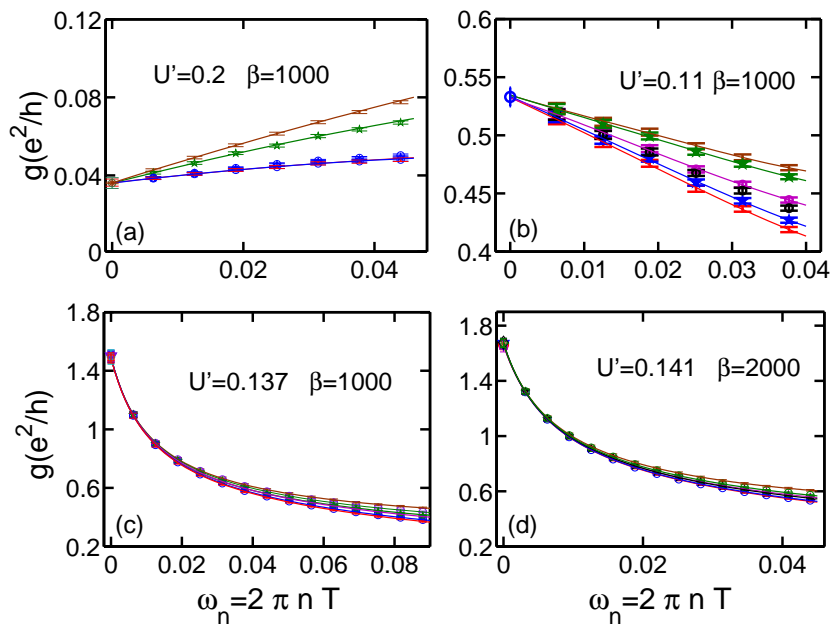}
\vspace*{-0.2in}
\caption{(color online) Conductance at Matsubara frequencies (symbols) for
quadruple dot system and the corresponding fits used to extrapolate to zero
frequency (lines). In the calculation, the tunneling $t=0.3$ and the coupling
$\Gamma=0.2$. The values of $U'$ and $\beta$ are (a) $0.2$, $1000$, (b) $0.11$,
$1000$; (c) $0.137$, $1000$; (d) $0.141$, $2000$.
}
\label{fig:cond_QMC}
\end{figure}

In this section, we show how to obtain the conductance from quantum Monte Carlo
data by the extrapolation method; a longer description, as well as checks, is in
Ref.\,\cite{dong10a}. The quadruple quantum dot system can be mapped to a
one-dimensional infinite tight-binding chain as shown in Fig.\,\ref{fig:chain},
where a pseudospin is considered on each site (corresponding to $+$ or $-$ in
Fig.\,1 of the main text). The linear conductance is obtained \cite{Syljuasen07a,
dong10a} by extrapolating the conductance at the (imaginary) Matsubara
frequencies, $g(i\omega_n)$ with $\omega_n\!=\!2\pi n T$, to zero frequency,
$G\!=\!\lim_{\omega_n\rightarrow 0}g(i\omega_n)$, 
\begin{equation}\label{eq:giwn}
 g(i \omega_n) = \frac{\omega_n}{\hbar} \int_{0}^{\beta} d\tau 
 \cos(\omega_n \tau)\langle P_x(\tau) P_y(0) \rangle \;,
\end{equation}
where $P_y$ is the sum of the electron charge density operators to the right of
$y$, $P_y \equiv \sum_{y^{'}\geq y}\hat{n}_{y^{'}}$. Not all combinations of $x$
and $y$ can be used in Eq.\,(\ref{eq:giwn}) because the system is not a physical
chain, but only effectively mapped to a chain. Notice that the current through
the five bonds closest to the quantum dot sites (labeled $L0$ and $R0$)
correspond to the physical current. Therefore, $x$ and $y$ must be chosen from
among $\{L1, L0, R0, R1, R2\}$. In addition, left-right symmetry reduces the
number of independent combinations. In our calculation, we choose six cases for
$x$ and $y$: $(L0,L0)$, $(R0,R0)$, $(L1,L0)$, $(L1,R0)$, $(L1,R1)$, and
$(L0,R0)$. We carry out this extrapolation as in Refs.\,\onlinecite{Syljuasen07a}
and\;\onlinecite{dong10a}. For the CO and LSS regions, the first several QMC data
points [$g(i \omega_n)$, $n=1,\dots,m$, where $m=4$ to $6$ depending on the
case] can be fit to a quadratic polynomial, as in Fig.~\ref{fig:cond_QMC}(a)(b).
For the Kondo regime, the first $14$ QMC data points [$g(i \omega_n)$, $n=1
\dots 14$] are fit to a series of rational polynomial functions [see
Fig.\,\ref{fig:cond_QMC}(c)(d) for examples]. The conductance is the average of
the different extrapolation results, and the error bar corresponds to the
maximum spread. For all of the cases studied in this work, the extrapolation
appears to be straight forward and reliable (for cases in other systems for
which the extrapolation does not work well, see Ref.\,\onlinecite{dong10a}).

\section{II. Peak of Conductance near Resonance}

\begin{figure}[t]
\centering
\includegraphics[width=3.2in,clip]{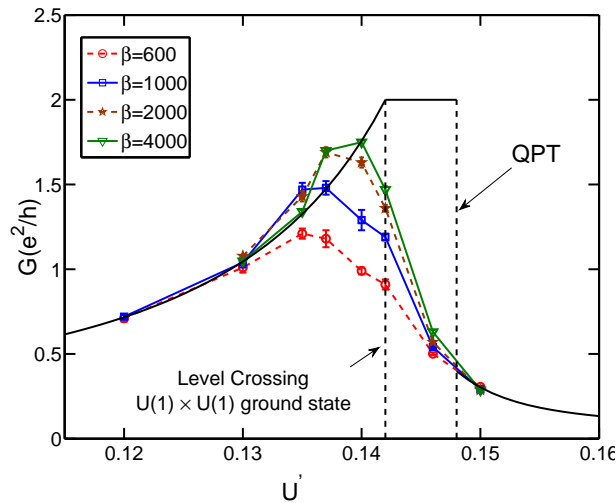}
\caption{(color online) Zero bais conductance as a function of $U'$ for
different $\beta$ in the regime of the peak caused by the $U(1) \!\times U(1)$
ground state. The two vertical dashed lines show the level-crossing and QPT
points. The black solid line gives the conductance at zero temperature
schematically. 
}
\label{fig:cond_zoomin}
\end{figure}

Fig.~\ref{fig:cond_zoomin} shows that the conductance peak approaches the
level-crossing point as the temperature approaches zero, as expected from the
$T=0$ theory, but that the peak is not at the level-crossing point at non-zero
$T$. This shift in the peak position can be understood as follows. When the
parameter $U'$ increases toward the level-crossing point $U_\text{LC}'\approx
0.142$, the singlet $\Singlet$ is the ground state of the dots, and the level
splitting $\Delta$ between $\Singlet$ and the doublet $\{\Tpp,\Tmm\}$ states
decreases to zero at $U_\text{LC}'$. Following the argument in
Ref.\,\onlinecite{GlazmanHouches05a} concerning the effect of a magnetic field on
the usual quantum dot Kondo effect, we expect that the effect of the level
splitting $\Delta$ at zero temperature should be similar to that of the
temperature at zero level splitting, though $G(\Delta/\Tk,T=0)$ and
$G(T/\Tk,\Delta=0)$ may have different universal forms. For finite $\Delta \ll T
$, $G$ can be approximated by $G(T/\Tk,\Delta=0)$ which saturates at $2e^2/h$ at
low temperature. On the other hand, at $T=0$, $G(T=0,\Delta) < 2e^2/h$ (because
the LSS has a substantial weight in the ground state). Therefore, the
conductance $G(T/\Tk,\Delta)$ may show a peak at non-zero $T$, and does so for,
e.g., $U'=0.135$ in Fig.~\ref{fig:cond_zoomin}. Similarly, for fixed $T\neq0$,
$G(\Delta/\Tk)$ may peak at non-zero $\Delta$. In summary, although the
conductance peaks at $\Delta=0$ for $T=0$, at non-zero temperature, $G(U')$ may
show a peak at $U'<0.142$ corresponding to finite level splitting $\Delta$.

\section{III. Effective Theory Near the Level Crossing}

The $S$ operators defined in Eq.\,(5) of the main text satisfy the following
commutation relations:
\begin{align}
[S_+^{\rmnone},S_-^{\rmnone}]&=2S_z^{\rmnone}&
[S_z^{\rmnone},S_{\pm}^{\rmnone}]&=\pm S_{\pm}^{\rmnone}&\nonumber\\
[S_+^{\rmntwo},S_-^{\rmntwo}]&=2S_z^{\rmnone}&
[S_z^{\rmntwo},S_{\pm}^{\rmntwo}]&=\pm S_{\pm}^{\rmnone}&\nonumber\\
[S_+^{\rmnone},S_-^{\rmntwo}]&=2 S_z^{\rmntwo}&
[S_z^{\rmnone},S_{\pm}^{\rmntwo}]&=\pm S_{\pm}^{\rmntwo}&\\
[S_+^{\rmntwo},S_-^{\rmnone}]&=2 S_z^{\rmntwo}&
[S_z^{\rmntwo},S_{\pm}^{\rmnone}]&=\pm S_{\pm}^{\rmntwo}&\nonumber\\
[S_+^{\rmnone},S_+^{\rmntwo}]&=0&
[S_-^{\rmnone},S_-^{\rmntwo}]&=0&
[S_z^{\rmnone},S_{z}^{\rmntwo}]&=0 \;.&\nonumber
\end{align}
These relations generate the $SO(4)$ algebra, so the six $S$ operators form an
$SO(4)$ algebra. However, the six $M$ operators do \textit{not} form an $SO(4)$
algebra. If the standard basis for the fundamental representation of $SU(3)$ is
$F_{i}$, $i=1,2,\dots,8$ \cite{PeskinBooka}, the six $M$ operators can be written
as the linear combinations
\begin{align}
&M_z^{\rmnone}=F_3,\qquad\qquad\qquad\qquad\qquad\qquad\;
M_z^{\rmntwo}=(2/\sqrt{3}) F_8,\nonumber\\
&M_+^{\rmnone}=(F_4+iF_5-F_6+iF_7)/\sqrt{2},\qquad 
M_-^{\rmnone}=(F_4-iF_5-F_6-iF_7)/\sqrt{2},\\ 
&M_+^{\rmntwo}=(F_4+iF_5+F_6-iF_7)/\sqrt{2},\qquad 
M_-^{\rmntwo}=(F_4-iF_5+F_6+iF_7)/\sqrt{2} \;. \nonumber
\end{align}
The two missing operators $\Tpp\langle -\! -\! |$ and $\Tmm\langle +\! +\! |$
can be written as 
\begin{align}
\Tpp\langle -\! -\!| =F_1+iF_2,\qquad
\Tmm\langle +\! +\!| =F_1-iF_2 \;.
\end{align}
Therefore, the six $M$ operators combined with the two missing operators form an
$SU(3)$ algebra.

Up to order $\Gamma/U$, there is no direct process which leads to $\Tpp
\Leftrightarrow \Tmm$. The only path leading to $\Tpp \Leftrightarrow \Tmm$ is
via the singlet state as an intermediary: $\Tpp \Leftrightarrow \Singlet
\Leftrightarrow \Tmm$. When higher-order terms $(\Gamma/U)^{2}$ are considered,
the four electron hopping terms do produce a direct  $\Tpp \Leftrightarrow \Tmm$
process:
\begin{equation}
 \Hflip=A \sum_{k,k^{'},q,q^{'}}\left(\Tpp \langle -\! -\! | 
        c_{kL-}^{+}c_{k^{'}L+} c_{qR-}^{+}c_{q^{'}R+} + \text{h.c.}\right) \;.
\label{eq:Heff}
\end{equation}
Since $\Hflip$ is comprised of four electron operators, its naive scaling
dimension is negative, suggesting that $\Hflip$ is irrelevant. To check this,
consider the one-loop RG equations. The scaling equations for the system
combining $\Hflip$ with $H_{\rm Kondo}^{\rm eff} $ [Eq.\,(3) of the main text]
consist of the original four equations in Eq.\,(6) of the main text plus one
additional equation:
\begin{equation}
\frac{dA}{d\ln\! D} = -2 \rho J_{\perp}^{\rmnone} A \;.
\end{equation}
Solving these five equations numerically with $J_{\perp}^{\rmnone}>0$, we find
that the coupling $A$ flows to 0 for any initial value. Therefore, $\Hflip$ is
an irrelevant operator in the strong coupling phase, which confirms the naive
scaling-dimension analysis.

\section{IV. Effective Theory Near the QPT}

To study the physics near the KT quantum phase transition, we develop a low
energy effective theory closely following Refs.\,\onlinecite{Garst04a} and
\onlinecite{Galpin0506a}. First, consider the effective Hamiltonian in the large
$U'$ limit without tunneling, and note that the energy of $\Singlet$ is much
higher than that of $\{\Tpp,\Tmm\}$. Using $\Gamma/U$ as a small parameter, we
make a Schrieffer-Wolff transformation to integrate out $\Tzero$ and $\Singlet$;
higher-order terms are neglected. The resulting fixed point Hamiltonian is 
\begin{equation}
\HCO= 
\sum_{k, s=L/R,\sigma =+/-}\epsilon_{ks\sigma} c_{ks\sigma}^{+} c_{ks\sigma} +
K \sum_{k k^{'}s\sigma} \Big( \hat{n}_{L\sigma}+ \hat{n}_{R\sigma}-1 \Big) 
c_{ks\sigma}^{+} c_{k^{'}s\sigma} \;.
\end{equation}
The ground state corresponds to a \textit{charge ordered state} with two-fold
degeneracy, $\{ \Tpp, \Tmm \}$: the two electrons are frozen in either the upper
two quantum dots or the lower dots. The charge order can be screened by $\Hflip$
given in Eq.\,(\ref{eq:Heff}).
The model here is very similar to that for the orthogonality catastrophe in the
x-ray edge problem \cite{Anderson67a,MahanBooka}, as pointed out in
Refs.\,\onlinecite{Garst04a} and \onlinecite{Galpin0506a}; we briefly summarize
their argument here. 

The charge ordered state can be flipped by the operator 
$\hat{f}\equiv
\Tpp\langle - -\!|\, c_{kL-}^{+}c_{k^{'}L+} c_{qR-}^{+}c_{q^{'}R+}$. 
According to Hopfield's rule of thumb \cite{Hopfield69a}, the correlation
function in the charge-ordered phase is given by 
\begin{equation}
 \left\langle \hat{f}^{+}(t) \hat{f}(0)\right\rangle_{\HCO} \sim t^{-\alpha}
% \end{equation}
% where
% \begin{equation}
 \qquad \text{where} \qquad
 \alpha = \displaystyle\sum_{i=L+,L-,R+,R-} (\Delta n_{i})^2
 \label{eq:fcor}
\end{equation}
is related to the change in occupation $\Delta n_i$ of each dot. $\Delta n_{i}$
can be expressed in terms of the conduction band phase shift $\delta$ through
the Friedel sum rule; therefore, the anomalous exponet can be related to
$\delta$,
\begin{equation}
 \alpha = \sum_{i=L+,L-,R+,R-} (\Delta n_{i})^2=
 4\left(\frac{2\delta}{\pi}-1 \right)^2
 \;.
\end{equation}
In a 1D problem such as ours, the power-law decay of correlations in
Eq.\,(\ref{eq:fcor}) leads to the criteria
\[
\left\{ 
\begin{array}{l l l}
  \alpha/2 > 1 & \quad \text{irrelevant} & \quad \text{Charge Ordered state}\\
  \alpha/2 < 1 & \quad \text{relevant} & \quad \text{Charge Kondo state}\\
  \alpha/2 = 1 & \quad \text{marginal} & \quad \text{Critical Point}\\
\end{array} \right.
\]
because of the possible infrared (long time) singularity. Thus the criterion for
the critical point separating the charge-ordered and charge-Kondo states is
 \begin{equation}
  2\left(\frac{2\delta_{c}}{\pi}-1 \right)^{2}=1 \ \ \Longrightarrow \ \ 
  \delta_{c}=\frac{\pi}{2} \Big(1 - \frac{1}{\sqrt{2}} \Big)
  \;.
 \end{equation}
The phase shift $\delta$ depends on the coefficient $K$ of the potential
scattering term in $\HCO$, while $K$ itself depends on $U$, $U'$, and $\Gamma$.
Therefore, the phase shift is a function
of $U$, $U'$, and $\Gamma$, and
\begin{equation}
\delta_{c}(U,\Ukt,\Gamma)=\frac{\pi}{2} \Big(1 - \frac{1}{\sqrt{2}} \Big)
\qquad \Longrightarrow \qquad \Ukt=f(U,\Gamma)
\;.
\end{equation}
So, when $U>\Ukt$, we have $\alpha>2$, $\Hflip$ is irrelevant, and the system is
in the charge-ordered phase with a two-fold degenerate ground state. However,
when $U<\Ukt$, $\alpha<2$, $\Hflip$ is relevant, and so this four electron
hopping operator prodcues $ \Tpp \leftrightarrow \Tmm $ which screens the
charge-ordered state to form the charge-Kondo state. There is a KT-type quantum
phase transition \cite{Garst04a,Galpin0506a} at $ U' =\Ukt$.

In contrast to Refs.\,\onlinecite{Garst04a} and \onlinecite{Galpin0506a}, we must
consider the influence of the tunneling $t$ on the QPT (i.e.\ direct tunneling
between $\Lp$ and $\Rp$, and between $\Lm$ and $\Rm$). To include tunneling,
processes of order $\Gamma t/U^2$ need to be considered \cite{FN_SWhigherordera}.
The low energy effective Hamiltonian in the charge-ordered phase becomes
\begin{align}
 \HCOt &= 
 \sum_{k, s=L/R,\sigma =+/-}\epsilon_{ks\sigma} c_{ks\sigma}^{+} c_{ks\sigma} +
    K \sum_{k k^{'}s\sigma} 
    (\hat{n}_{L\sigma}+\hat{n}_{R\sigma}-1) c_{ks\sigma}^{+} c_{k^{'}s\sigma}
    \nonumber\\
&\quad +\widetilde{K_{t}} \sum_{k,k^{'},\ \sigma}
(\hat{n}_{L\sigma}+\hat{n}_{R\sigma}-1) 
(c_{kL\sigma}^+c_{k^{'}R\sigma}+c_{kR\sigma}^+c_{k^{'}L\sigma}) \;.
\end{align}
The contribution of the tunneling merely adds the possibility of potential
scattering between the electrons in the left and right leads; note that the
dependence on the filling of the four dots, either $\Tpp$ or $\Tmm$, is the
same. Thus, tunneling does not introduce a charge order flip process, but does
contribute to the phase shift experienced by the lead electrons when the dot
flips. We expect, then, that $t$ does not affect the essential physics of the
charge-ordered phase and QPT. We should check, however, whether the tunneling
leads to a relevant process in addition to $\Hflip$ which can screen the charge
order. The lowest order screening terms induced by tunneling are
 \begin{align}
 \hat{f}_{1}=|++\rangle\langle --| 
   c_{kL-}^{+}c_{k^{'}L+} c_{qR-}^{+}c_{q^{'}L+}\;, \qquad
 \hat{f}_{2}=|++\rangle\langle --|
   c_{kL-}^{+}c_{k^{'}L+} c_{qL-}^{+}c_{q^{'}L+} \;.
 \end{align}
Following the arguments given above, we find that the anomalous exponent for
$\hat{f}_1$ is $\alpha/2=2(2\delta/\pi-1)^{2} + 1 \geq 1$, while that for
$\hat{f}_2$ is $\alpha/2=2(2\delta/\pi-1)^{2} + 2 > 1$. Therefore, both
$\hat{f}_1$ and $\hat{f}_2$ are irrelevant operators in the charge ordered
phase. The tunneling does not affect the essential physics (i.e.\ the quantum
phase transition) of the system. Its influence is only felt through the phase
shift $\delta$ which depends on the tunneling $t$. $\Ukt$, then, is a function
of $U$, $\Gamma$, and now $t$. Thus, the tunneling $t$ \textit{does} affect
where the QPT occurs (Fig.\,1 in the main text); in particular, to observe the
QPT in experiments, $t$ can be tuned instead of or in addition to the
interaction $U'$. 

\vspace*{-0.2in}

%\bibliography{quadruple,cond_qmc,footnotes,supp}

\begin{thebibliography}{22}
\bibitem[{FN_({\natexlab{b}})}]{FN_SWhigherorder}
\bibinfo{note}{Processes of order $\Gamma t^n/U^{n+1}$ produce the same terms
  in the effective Hamiltonian as $\Gamma/U$ and $\Gamma t/U^{2}$ proceeses,
  and so do not lead to any essential difference.}
\end{thebibliography}

\end{document}